\documentclass[12pt,preprint]{aastex}

\def\cm2{cm$^{-2}$}

\def\lya{Ly$\alpha$}

\def\mgtwo{\ion{Mg}{2}}
\def\nhi{$N$(H~I)}

\shorttitle{Intermediate Mg~II Systems in CORALS}
\shortauthors{Ellison et al.}

\begin{document}


\title{The Number Density of $0.6 < z < 1.7$ Mg~II Systems from CORALS: 
Observational Biases at Intermediate Redshift\footnote{These data were 
obtained from the 3.6-m on La Silla
(70.A-0006), UT3 on Paranal (69.A-0053,71.A-0011), the WHT on La 
Palma (W/2002A/10) and the Baade telescope at Las Campanas}.}


\author{Sara L. Ellison}
\affil{University of Victoria, Dept. Physics \& Astronomy, 
Elliott Building, 3800 Finnerty Rd, Victoria, V8P 1A1, British Columbia,
Canada}
\email{sarae@uvic.ca}

\author{Chris W. Churchill}
\affil{Department of Astronomy,  New Mexico State University, 1320 Frenger 
Mall, Las Cruces, New Mexico, USA}
\email{cwc@nmsu.edu}

\author{Samantha A. Rix\altaffilmark{2}, Max Pettini}
\affil{Institute of Astronomy, Cambridge, Madingley Rd, Cambridge, CB3 0HA, UK}
\email{srix@ing.iac.es,pettini@ast.cam.ac.uk}

\altaffiltext{2}{Current address: Isaac Newton Group, Apartado 321, 
38700 Santa Cruz de La Palma, Spain}

\begin{abstract}
The goal of the Complete Optical and Radio Absorption Line System (CORALS) 
survey is to quantify the potential impact on QSO absorber statistics
from dust in intervening galaxies.
Dust may introduce a selection bias in surveys which are
based on magnitude limited QSO samples, leading to an underestimate
of absorber number density, $n(z)$.  Here we present the results of the second 
phase of the CORALS survey which 
extends our previous work on $z > 1.8$ damped Lyman $\alpha$ systems 
(DLAs) to search for strong metal line systems (candidate DLAs) in
the range 0.6 $< z <$ 1.7.  We have identified 47 Mg~II systems 
with rest frame equivalent widths $EW$(\mgtwo $\lambda 2796) > 0.3$ \AA\
in our sample of 75 radio-selected quasars.
The total redshift path covered by the survey is $\Delta z$ = 35.2, 58.2
and 63.8 for $EW$(\mgtwo $\lambda 2796) >$ 0.3, 0.6 and 1.0 \AA\ thresholds
respectively
(5 $\sigma$).  Our principal and most robust result is that the $n(z)$ of 
low redshift \mgtwo\ systems
determined for the CORALS survey is in excellent agreement with that
of optically-selected, magnitude limited QSO samples.  
We use empirically determined \mgtwo\ equivalent width statistics to 
estimate the likely number of DLAs in this sample.  The statistically
inferred number density of DLAs, $n(z)=0.16^{+0.08}_{-0.06}$, is 
consistent with other low redshift samples, although the
large 1$\sigma$ error bars permit up to a factor of 2.5 more DLAs
in CORALS.    However, confirmation of the DLA candidates, precise
evaluation of their $n(z)$ and measurement of their
H~I column densities awaits UV observations
with the \textit{Hubble Space Telescope}.    
Finally, we report an excess of intermediate redshift \mgtwo\ systems
observed towards bright QSOs which could be due to a lensing
amplification bias.  However, there is also evidence that this
excess could simply be due to more sensitive EW detection limits towards
brighter QSOs.  We also emphasize that absorber statistics 
determined from magnitude limited surveys reach a steady value if the
completeness limit is significantly fainter than the fiducial value of the 
quasar luminosity function.
\end{abstract}





\section{Introduction}

Modern instrumentation has provided us with two techniques capable
of identifying high redshift ($z \gtrsim 3$) galaxies in relatively
large numbers.  Direct detection of star-forming galaxies
via deep multi-color imaging employs broad band photometry
in filters which span characteristic continuum breaks 
(e.g. Steidel et al. 1996; Lowenthal et al. 1997; Giavalisco 2002; 
Adelberger et al. 2004).
A complementary approach uses spectroscopy of distant QSOs whose 
lines of sight penetrate intervening gas clouds associated with
galaxies and the intergalactic medium
(e.g. Wolfe et al. 1986).  These two techniques occupy different 
astrophysical niches; the former identifies the brightest,
most actively star forming galaxies at early epochs (e.g.
Sawicki \& Yee 1998; Shapley et al. 2001; Nandra et al. 2002)
whereas the latter selects galaxies via their absorption cross
section of neutral hydrogen.  In particular, damped \lya\ systems (DLAs) 
are often proclaimed to be `unbiased' tracers 
of galaxy evolution and are expected to provide a fair census of
neutral gas at all redshifts.
However, despite the lack of Malmquist bias, DLA statistics are
susceptible to other possible selection effects.
For example, dust in intervening galaxies may cause significant
extinction of background sources which would make dusty absorbers 
more difficult to detect in optical surveys.  
Obfuscation of background QSOs is also theoretically supported 
(Ostriker \& Heisler 1984; Fall \& Pei 1993; Masci \& Webster 1995 )
and often invoked as an explanation for observed trends
such as the anti-correlation between the neutral hydrogen column density,
$N$(\ion{H}{1}), and metallicity (e.g. Prantzos \& Boissier 2000). 
However, Murphy \& Liske (2004) have recently determined $E(B-V)$
values of less than 0.01 magnitudes (assuming an SMC extinction curve) 
for DLAs in the Sloan Digital Sky Survey.
Selection against evolved galaxies may also be
responsible for the lack of metallicity evolution (Pettini et al. 1999)
in DLAs.  If it is shown that dust obscuration is not an
important effect, then we
must appeal to alternative explanations for these observed trends.

In order to specifically address the issue of dust biasing in
DLA surveys, we have designed the Complete Optical and Radio Absorption 
Line System survey (CORALS).  CORALS is based on
a sample of radio-selected quasars (Jackson et al. 2002) 
with complete optical identifications. The key here is that
QSO selection was executed at a wavelength immune to extinction,
and yet optical counterparts have been identified for every target.
The high redshift ($z_{\rm abs}>1.8$) results of CORALS~I 
(Ellison et al. 2001) indicate that the effect of dust bias 
at this redshift is relatively minor, although a larger 
radio selected sample would be desirable to improve
the statistics.  These results do not indicate that dust is 
entirely absent
in high redshift DLAs.  On the contrary, there is evidence
from a number of observations including chemical abundances,
(Pettini et al. 1997), QSO colors and spectral indices
(Pei, Fall \& Bechtold 1991; Outram et al. 2001) and Lyman
break galaxies (Sawicki \& Yee 1998; Shapley et al. 2003) that dust is
present at these early times.  Apparently extinction
in the majority of gas-rich galaxies at $z>2$ has only a minor effect
on the statistics of QSO surveys with optical magnitude limits
$V \lesssim 20$.

At low redshift, however, the story may be quite different.  The
most prominent sources of dust today are the envelopes of
evolved, cool, stars which can be distributed into the ISM by winds.
However, in order to explain the large amounts of dust already
present at high redshift (e.g. Priddey \& McMahon 2001; Bertoldi et al. 2003), 
additional sources such
as supernovae (Morgan \& Edmunds 2003; Dunne et al. 2003) and AGN 
(Elvis, Marengo \& Karovska 2002) have been proposed,
although our understanding of dust formation at any epoch remains
sketchy.  Regardless of the source of dust, it seems plausible that
stars (and possibly AGN) must be well established before grains
can be distributed in the ISM.  The epoch around $z=2$ -- 3 is
emerging as an important formative stage in the universe's history,
with both star formation and QSO space density declining steeply after
this time (Madau et al. 1996; Shaver et al. 1996).  Moreover,  
Dickinson et al. (2003) and Rudnick et al. (2003) have shown that 
this is the epoch when stellar mass sees marked
evolution; only about 5 -- 10\% of galactic stellar mass is 
apparently in place before a redshift of 2.5, but 50 -- 75\% at $z \sim 1$.  
Models of chemical evolution also predict a more serious dust bias
at low redshift (e.g. Masci \& Webster 1999; Churches, Nelson \& Edmunds 
2004).  Moreover,
our only direct evidence of spectral dust features seen in absorption is
at $z < 1$.    Malhotra (1997) found evidence for
the 2175 \AA\ absorption feature in a composite of 96 \mgtwo\ systems
with an average redshift of $z\sim 1.2$.  The same feature has also
recently been identified in a $z \sim 0.5$ DLA (Junkkarinen
et al. 2004) and in a $z =0.8$ lensing galaxy (Motta et al. 2002).
No evidence for this extinction feature has been found at higher
redshift (Fall, Pei \& McMahon 1989).  This clearly highlights the 
need to quantify the dust bias at $ z< 1.5$.

In order to extend the CORALS~I survey to lower redshifts
where dust bias may be more pronounced, we have executed a complementary
follow-up survey.  In this paper, we present the results of
this survey which targets \ion{Mg}{2} absorption systems primarily
in the range $0.6 < z < 1.7$,
once again utilizing an optically complete radio-selected sample
of QSOs.  A survey for strong \ion{Mg}{2} systems is a stepping
stone towards the quantification of dust bias in DLAs, which
form a subset of the systems discovered here.  Future follow-up observations
with the Hubble Space Telescope (\textit{HST}) 
will allow us to confirm the \ion{H}{1} column
densities of the candidate DLAs discovered by CORALS and determine
unbiased statistics for this population at intermediate redshifts.
We describe the design of the survey
(\S2), the observations (\S3) and tabulate the absorption systems 
identified (\S4).  We present our absorption line statistics in
\S5 and \S6.  Discussion and conclusions can be found in 
\S7 and \S8 respectively.

\section{Survey Design}

The original CORALS~I sample selected all QSOs with $z_{\rm em} > 2.2$ from
the Parkes 0.25~Jy flat spectrum sample of Jackson et al. (2002).
We imposed this lower redshift bound because our survey was limited
to DLAs at  $z_{\rm abs} > 1.8$. For DLAs at lower redshifts, the \lya\
line occurs at a wavelength where the throughput 
of most spectrographs drops significantly and at $z_{\rm abs} \lesssim 1.6$
\lya\ falls bluewards of the atmospheric cut-off, completely precluding 
ground-based detection. 
Since \lya\ observations with the \textit{HST} 
for an optically complete sample are not feasible due to its limited aperture, 
we adopted an alternative
strategy for our low redshift CORALS survey (CORALS~II).  Based on 
the approach of Rao \& Turnshek (2000, RT00), we selected systems with
large rest equivalent widths (EWs) of \mgtwo\ $\lambda\lambda 2796, 2803$ 
and \ion{Fe}{2} $\lambda 2600$ as candidate DLAs.
These metal lines can be efficiently observed in ground-based optical
spectra over the range $0.6 \lesssim z \lesssim 1.8$.  
We can assess the severity of dust bias in previous surveys by
determining the number of \mgtwo\ systems, as well as the statistically
inferred number of DLAs.

In order to extend the CORALS survey to lower
redshifts using \mgtwo, it is necessary to re-define the original
QSO sample so that it is optimized for this task.  Moreover, the
observations from CORALS~I are not suitable for the identification of
moderate redshift \mgtwo\ for two reasons.  First, we must target the 
continuum 
redward of \lya\ whereas the wavelength range of CORALS~I spectra often
covers only the \lya\ forest.  Second, the CORALS~I spectra were 
of insufficient resolution and
signal-to-noise ratio (S/N) for consistent identification of 
the \mgtwo\ doublet (a minimum resolving power 
$R \equiv \lambda/\Delta \lambda \simeq 900$ is required in order to
resolve the Mg~II $\lambda \lambda 2796, 2803$ doublet).
Thus, for the present CORALS~II survey we have
selected all sources with $1.80 < z_{\rm em} < 2.55$
from the parent sample of 0.25 Jy Parkes flat spectrum QSOs.
The high systemic redshift cut-off was selected so that
even for the lowest redshift Mg~II systems (typically $z_{\rm abs} \sim 0.6$)
the \mgtwo\ $\lambda\lambda$2796, 2803 \AA\ doublet was always 
located redwards of \lya\
emission.  Although the corresponding lower systemic redshift cut-off 
could have consequently extended to $z_{\rm em}=0.6$, this would have yielded
an unmanageble number of survey sightlines given observational
facilities at our disposal.  The lower redshift cut-off
was therefore set to $z_{\rm em}=1.80$, resulting in a sample of 75 QSOs.

\section{Observations and Data Reduction}

Spectra were obtained with a variety of instruments on 4- to 8-m telescopes:
ISIS (William Herschel Telescope), EFOSC2 (ESO 3.6-m), the Boller \&
Chivens (B\&C; Magellan-Baade) and FORS1 (VLT UT3).  All observations
executed at the VLT were obtained in service mode.
A summary of the observation dates and instrumental setups is given in Table 
\ref{obs_table}.
In Table \ref{qso_table} we list the observed QSOs, exposure times,
magnitudes and redshifts.  
The transparency conditions were generally clear to photometric and
seeing ranged from approximately 0.5 to 1.5 arcsec.  For EFOSC2, B\&C
and ISIS, the slit was aligned at the parallactic angle at the start
of each observation; for FORS1 we used the atmospheric dispersion corrector.

All but two of the 75 QSOs in our sample have been observed, availability
of telescope time precluding the observation of two of our faintest
targets.  Although the main objective of this survey is to achieve
optical completeness,
the exclusion of two sightlines is extremely unlikely to affect 
our number density statistics.

\begin{figure}
\centerline{\rotatebox{270}{\resizebox{12cm}{!}
{\includegraphics{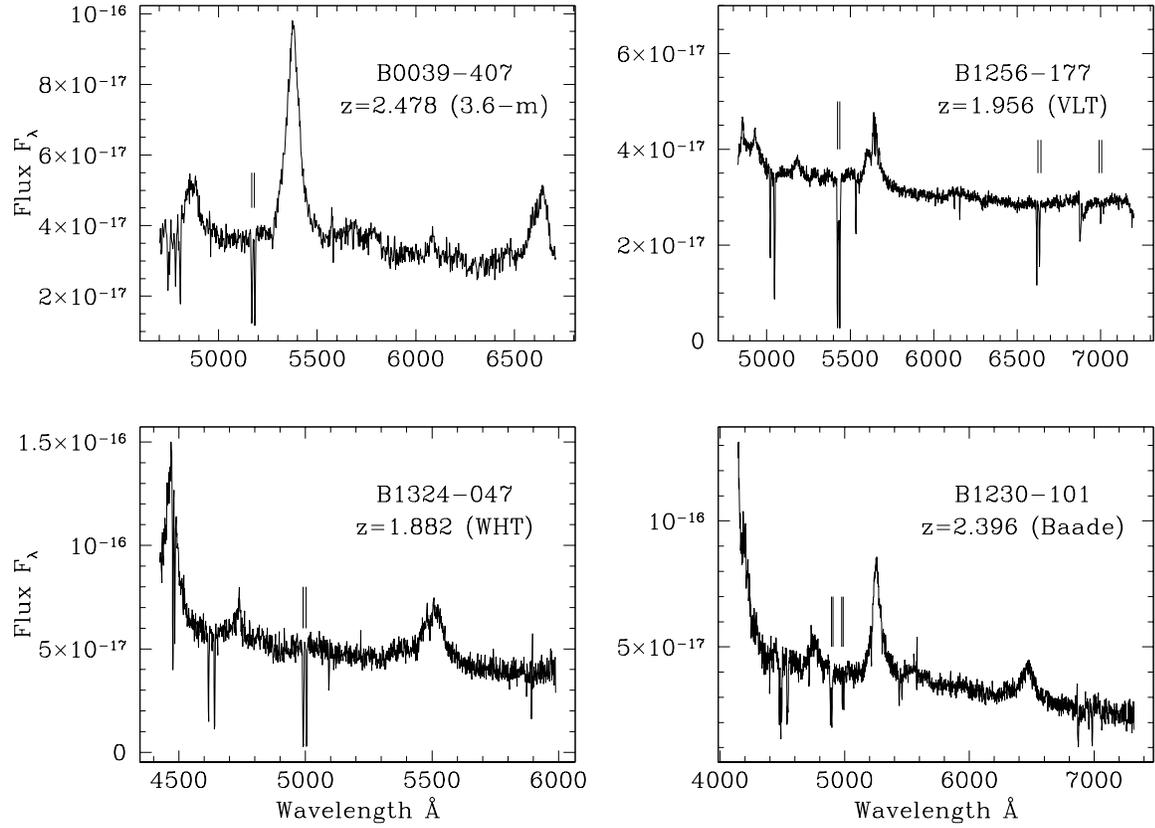}}}}
\caption{\label{spec_fig} Examples of spectra obtained with the four telescopes
utilized for this program.  \mgtwo\ systems listed in Table \ref{line_table}
are indicated with vertical tick marks.}
\end{figure}

The data were reduced using standard IRAF\footnote{IRAF is written and 
supported by the IRAF programming group at the National Optical 
Astronomy Observatories.} routines to execute the usual
corrections for flat fields and bias structure.  We note that
EFOSC2 has a very narrow overscan strip ($\sim$ 6 pixels) that
could not be used, so that only the average bias frame was subtracted.
Special attention was paid to the flat fielding of the ISIS spectra
where significant structure is introduced by the dichroic response,
by vignetting on the new large format red arm CCD (applicable to the 
October 2002 data), and by interference patterns at red wavelengths.  
The spectra were optimally extracted and wavelength calibrated using
HeAr (EFOSC2), CuAr+CuNe (ISIS), HeArNe (B\&C) or HeNeArCdHg (FORS1) lamps.  
In the case of FORS1 and EFOSC2, the instrument flexure is minimal 
and arcs were obtained only at the start and the end of each night.  
For ISIS and the B\&C spectrographs, arcs were
taken before or after each science target.  Although the observing conditions
did not allow precise absolute flux determination, 
the intrinsic continuum shape has been recovered by flux calibrating
with spectrophotometric standards taken through the night.  The
calibrated spectra were then combined and shifted to a vacuum
heliocentric reference frame.  Finally, a correction for Galactic
interstellar reddening was applied assuming the empirical selective
extinction function of Cardelli, Clayton \& Mathis (1989) and
E(B$-$V) values taken from the DIRBE extinction maps of 
Schlegel, Finkbeiner \& Davis (1998) with an assumed $R_V$=3.1.
In Figure \ref{spec_fig} we present examples of the spectra obtained
for this project from each telescope utilized.

\section{Mg~II System Doublet Identification}

The search for \mgtwo\ $\lambda\lambda$ 2796, 2803 \AA\
doublets was undertaken using an automated
search algorithm based on the technique 
originally designed for identification of absorption lines in 
the $HST$ QSO key project
spectra (Schneider et al. 1993).  We have refined this technique;
full details can be found in Churchill et al (2000a).
Here we give only a brief qualitative summary of the steps involved.

For each flux spectrum, we calculated an EW and an EW uncertainty spectrum. The
EW uncertainty spectrum provides a running detection limit as a function of
wavelength (e.g., Schneider et al. 1993; Churchill et al. 1999).  
These spectra are
produced by assuming an unresolved line centered on each pixel with weighting
of neighbouring pixels by the appropriate instrumental spread function. Using
these spectra, we ran an automated routine to
objectively locate candidate \mgtwo\ doublets.  
During this automated process, a preliminary measurement of
the equivalent widths and the \mgtwo\ doublet ratio is computed, again assuming
unresolved absorption features.  

The objectively determined candidate doublet list is then visually inspected
using the flux spectra to ascertain which are \textit{bona fide} 
\mgtwo\ doublets.  Many
candidates were corroborated by the detection of
one or several \ion{Fe}{2} lines and the \ion{Mg}{1}
$\lambda 2853$ line. Following visual inspection, roughly 10\% of the 
candidates
were discarded.  We also visually inspected the flux spectra to search for
\mgtwo\ candidates that may have been missed in the automated search.  We found
no candidates missed by the search algorithm.

We then interactively fitted Gaussian profiles to the absorption lines using a
$\chi^2$ minization routine to determine the final EWs and redshifts (line
centers), and their uncertainties (Churchill et al. 2000a). For unresolved 
lines,
the Gaussian width was automatically held at the instrumental spread function
width.  In the few cases of highly resolved lines, multiple Gaussians were used
to determine the EWs.   The systems were again inspected to confirm that the
redshifts of the $\lambda 2796$ and $\lambda 2803$ lines were consistent within
1~$\sigma$ and that they had physically meaningful doublet ratios within
the uncertainties.   For each system, we then performed Gaussian fitting on
statistically significant \ion{Mg}{1} and \ion{Fe}{2} transitions, or 
computed the
3~$\sigma$ EW limit for an unresolved feature from the EW uncertainty spectrum
at the expected location of the line.
The full list of \mgtwo\ systems and their \mgtwo\ $\lambda 2796$
and \ion{Fe}{2}~$\lambda 2600$~ EWs is given in Table \ref{line_table}.

\section{Survey Coverage}

\begin{figure}
\centerline{\rotatebox{270}{\resizebox{12cm}{!}
{\includegraphics{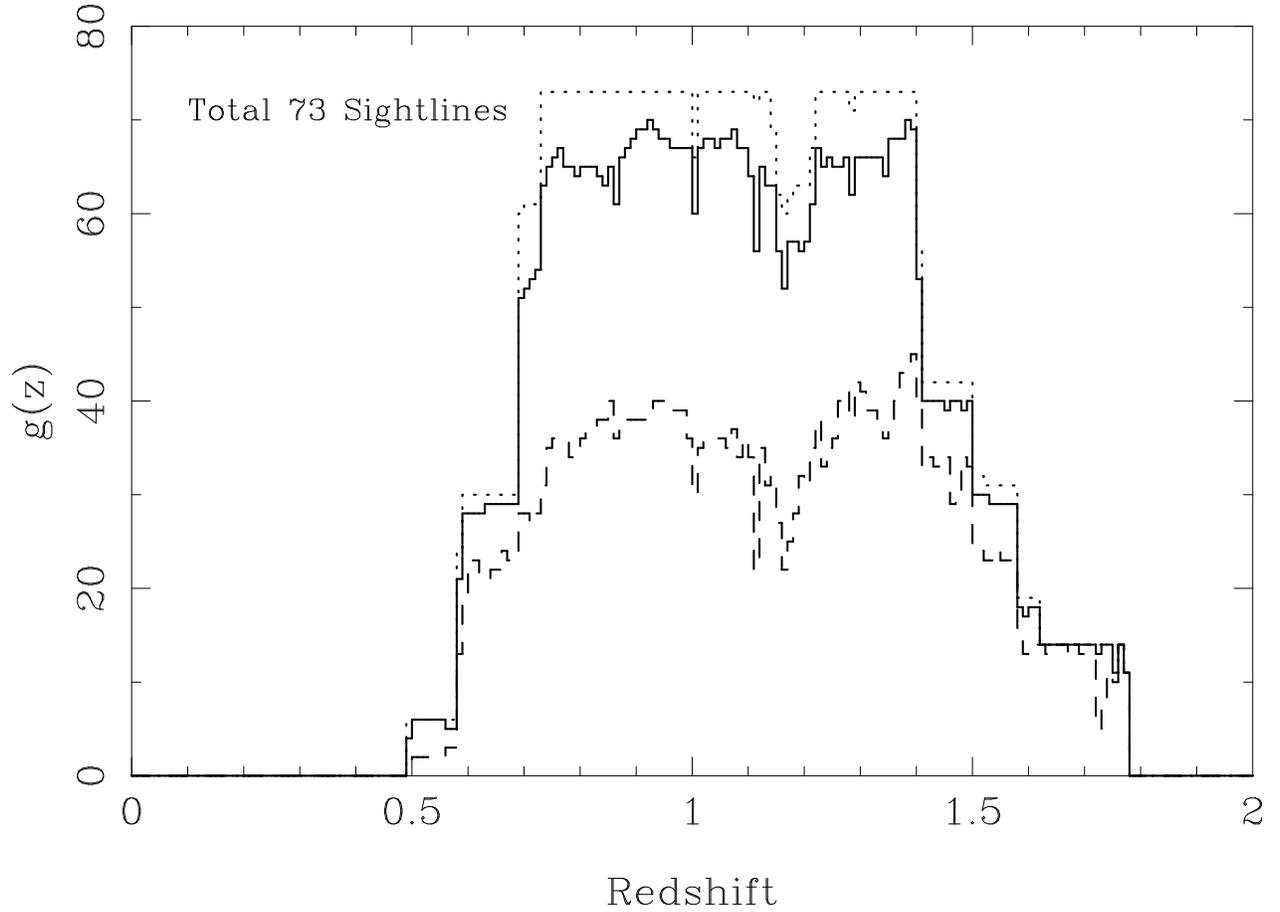}}}}
\caption{\label{gz}The redshift path density, $g(EW_{min},z_i)$ for three
minimum EW limits, as a function of redshift.  Dotted, solid and dashed
lines are 1.0, 0.6 and 0.3\AA\ (at 5 $\sigma$ significance) respectively. }
\end{figure}

The advantage of our line detection algorithm is that it computes the 
effective detection limit at each resolution element, allowing
us to easily calculate the survey coverage as a function of EW 
threshold.  The redshift path density for a given EW threshold, 
$g(EW_{min},z_i)$, is the number
of lines of sight in our survey at which a \mgtwo\ doublet of rest
equivalent width $EW_{min}$ at redshift $z_i$ could have been detected:

\begin{equation}
g(EW_{min},z_i)=\Sigma_j H(z_i - z_j^{min}) H(z_j^{max} - z_i) H[EW_{min}-w^j_{min}(z_i)]
\end{equation}

for $j$ QSOs in the survey.  $H$ is the Heaviside step function, 
$z^{min,max}$ correspond to the minimum and maximum redshifts at 
which \mgtwo\ were searched and $w^j_{min}(z_i)$ is the calculated
EW detection limit at $z_i$.  In Figure \ref{gz} we plot $g(EW_{min},z_i)$
for our survey, adopting three EW thresholds: $EW_{min} >$ 
0.3, 0.6 and 1.0 \AA\ (5 $\sigma$).
This figure illustrates that for the range $0.7 < z < 1.4$ we are
almost 100\% complete (except at $z\sim 1.18$ due to a gap in the
wavelength coverage of the ISIS spectra) for $EW_{min}$=0.6, 1.0 \AA\ 
and about 50\% complete for a threshold EW limit of 0.3 \AA.  

The total redshift path, $\Delta z$, of the survey is given by

\begin{equation}
\Delta z = \int_0^\infty g(EW_{min},z_i) dz
\end{equation}

\begin{figure}
\centerline{\rotatebox{270}{\resizebox{12cm}{!}
{\includegraphics{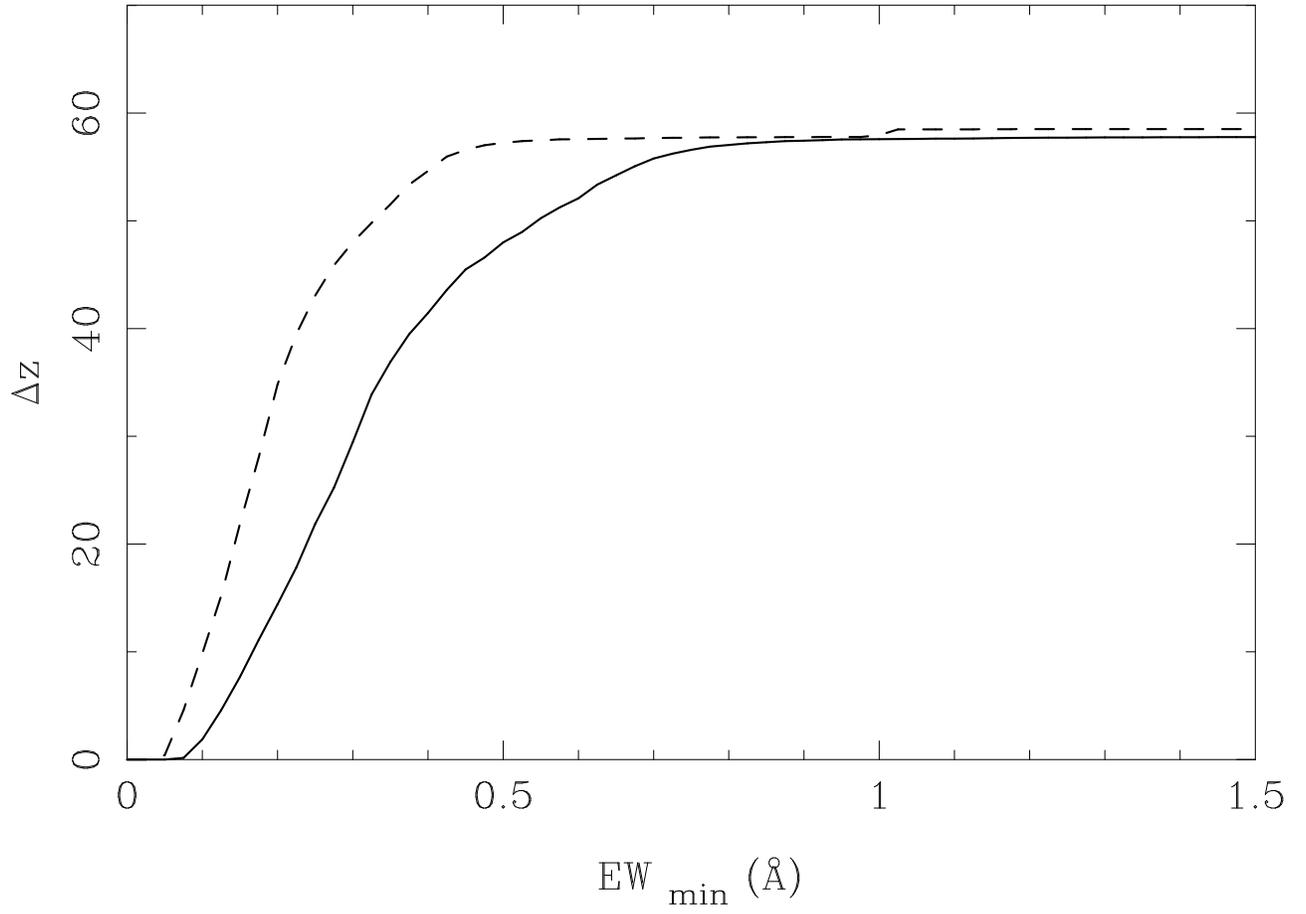}}}}
\caption{\label{dz}The cumulative redshift path ($\Delta z$) covered as 
a function 
of rest frame EW.  The solid line is for 5 $\sigma$, dashed is 3 $\sigma$.}
\end{figure}

In Figure \ref{dz} we plot the total redshift path as a function
of EW limit for both 3 and 5 $\sigma$ detection
limits. Corresponding values of $\Delta z$ are listed in Table \ref{dz_table}
for various combinations of $EW_{min}$ and detection 
significance as a supplement to the information in Figure \ref{dz}.
Table \ref{dz_table} additionally includes redshift coverage values
for simultaneous detection of \mgtwo\ $\lambda 2796$ and 
\ion{Fe}{2}~$\lambda 2600$; this point will be discussed in \S \ref{dla_sec}.  

\section{Results}

\subsection{Mg~II Number Density}

\begin{figure}
\centerline{\rotatebox{0}{\resizebox{12cm}{!}
{\includegraphics{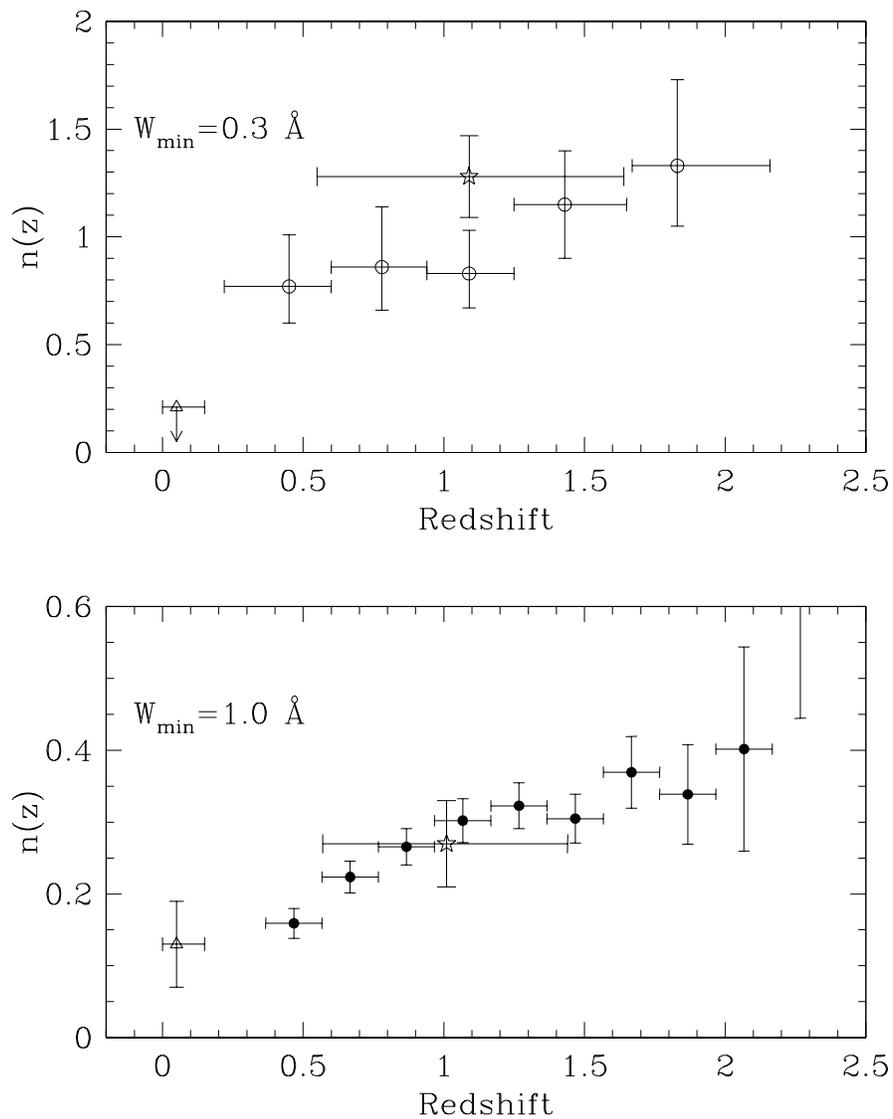}}}}
\caption{\label{nz}The number density of \ion{Mg}{2} systems, $n(z)$, in
the CORALS~II survey (open stars) compared with Steidel \& Sargent (1992),
preliminary results from the SDSS EDR (Nestor et al. 2003a) and 
Churchill (2001) which are plotted with 
open circles, solid circles and open triangles respectively.
Results are shown for two minimum EW thresholds. }
\end{figure}

The number density of absorbers
per unit redshift, $n(z)$, is computed for a given mean absorption redshift
by dividing the number of systems by $\Delta z$.
In Table \ref{stats} we present the values of $\Delta z$ and number of
absorbers for CORALS~II,
as well as the large surveys of Steidel \& Sargent (1992, SS92) 
and RT00.  Figure \ref{nz} compares 
the $n(z)$ determined for CORALS~II and other surveys as a function
of redshift.  We use the statistics of SS92 for $EW>0.3$ \AA\ and preliminary
statistics from the Sloan Digital Sky Survey (SDSS) 
Early Data Release ($EW> 1.0$ \AA, Nestor
et al. 2003a).  Since CORALS is a smaller survey than the
SDSS or SS92 and evolution in the number density is very mild over the
redshift range that we cover, we combine our $n(z)$ statistics into one 
redshift bin.  The results in Figure \ref{nz} demonstrate an excellent
agreement between the CORALS value of $n(z)$ and that determined from
the SDSS survey of Nestor et al.  The SDSS results for the intermediate
EW threshold of 0.6 \AA\ (Nestor et al. in preparation) are also
in excellent agreement with the CORALS value.
In the lower EW range, $EW>0.3$ \AA,
there is marginal evidence for a higher number density (at most a factor
of two) in the CORALS sample compared with SS92.
If this is interpreted within the framework of a dust bias, we would
conclude that the weakest \ion{Mg}{2} systems are more prone to extinction
than high EW systems.  This seems unlikely given that a) the lowest
EW \ion{Mg}{2} systems are likely to be associated with the outer regions
of galaxies (e.g. Churchill et al. 2000b; Ellison et al. 2004) and
b)  strong \ion{Mg}{2} systems statistically trace gas with higher
metallicities than weak systems (Nestor et al. 2003b).  
In any case,  Figure \ref{nz} illustrates
the main result of this survey: 
previous magnitude limited surveys have not under-estimated the number 
density of \ion{Mg}{2} systems in the range $0.6 < z < 1.6$.

\subsection{The Impact of Lensing Amplification Bias}\label{lens_sec}

Although the principal
focus of the CORALS survey has been the quantification
of dust bias, we consider here also the potential effect
of gravitational lensing.
The possibility that an amplification bias
can skew distant source counts has been discussed extensively
in the literature (e.g. Turner 1980; 
Schneider 1987; Hamana et al. 1997).  Amongst others,
Bartelmann \& Loeb (1996),
Smette et al. (1997) and Perna, Loeb \& Bartelmann (1997) have discussed
the specific case of amplification bias in QSO absorption system
surveys.  These papers describe the dual effect of magnification
of the background source and deflection of the sightline from the
central part of the lensing galaxy.
However, since the observational spotlight has focussed in large part
on the detection of \textit{high redshift} DLAs, little concern has been 
directed towards this possible bias, since lensing becomes less efficient as
the redshift of the intervening galaxy approaches that of the QSO.  
Only in the last few years have large numbers of low redshift DLAs been 
discovered and studied, renewing interest in the potential effect of the
amplification bias.

Le Brun et al. (2000) studied 7 QSOs with $z<1$ DLAs.  In no case 
did they find multiple images and they therefore concluded that 
amplification of the sources was at most 0.3 magnitudes.  
However, these authors also noted that these QSOs are fainter 
than those in many low redshift DLA surveys
and are therefore less likely to be affected by
amplification bias.  Using a large sample of 2dF QSOs, Menard \& 
P\'eroux (2003, MP03) have reported the first convincing evidence 
of amplification bias by showing that 
bright QSOs ($B \lesssim 19.5$) are more likely to 
have intervening \ion{Mg}{2} absorbers than fainter ones.  However,
P\'eroux et al. (2004) have also shown that gravitational lensing bias
has no demonstrable effect on the total column density of intervening
\ion{H}{1} gas and the mass fraction of neutral gas
in low redshift absorbers.

\begin{figure}
\centerline{\rotatebox{0}{\resizebox{12cm}{!}
{\includegraphics{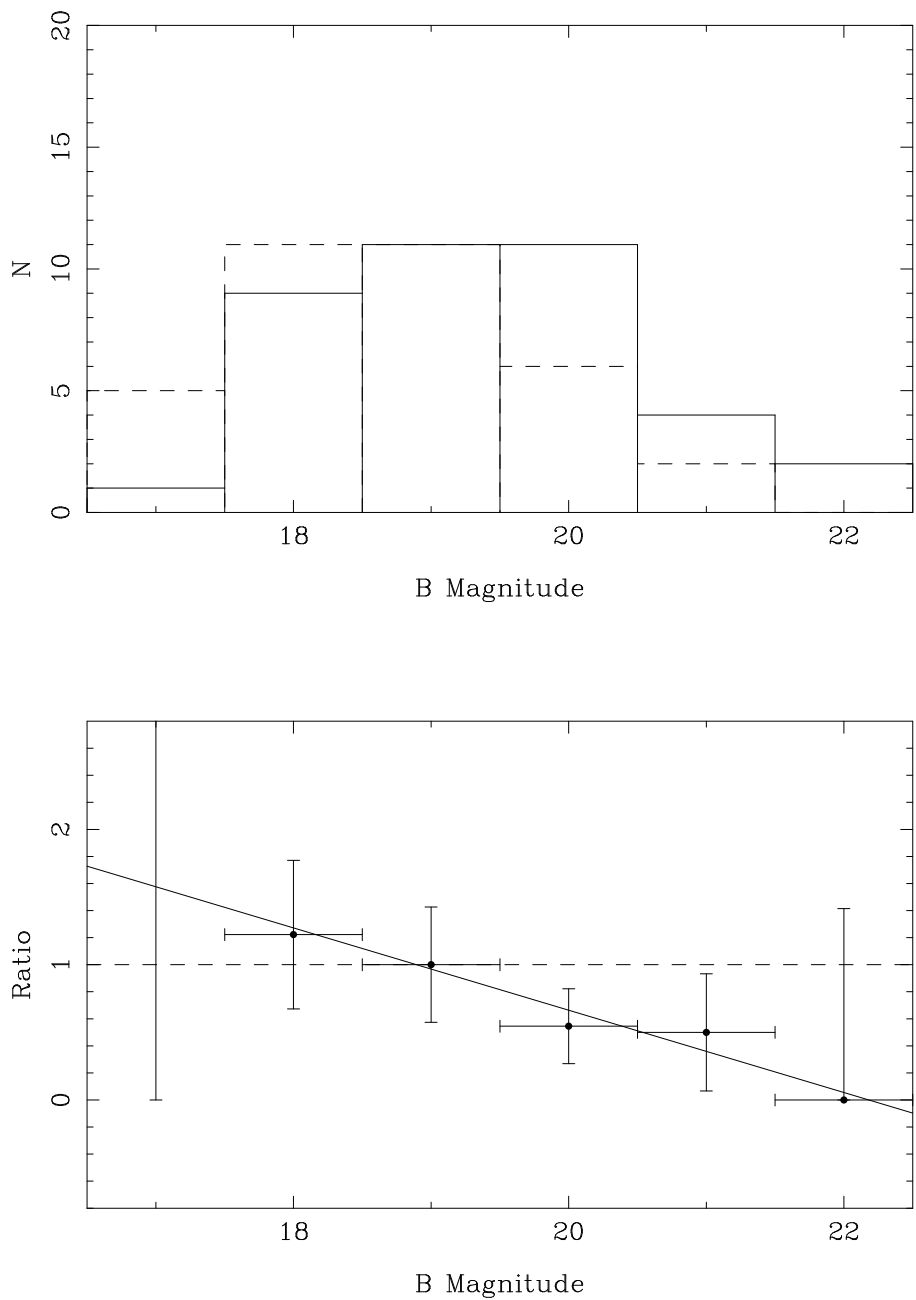}}}}
\caption{\label{mag_ratio}Top:  $B$-band magnitude distributions for QSOs
with (dashed line) and without (solid line) intervening \ion{Mg}{2} 
systems.  All \mgtwo\ systems in Table \ref{line_table} are
included.
Bottom:  The ratio of the number of QSOs with:without intervening
systems in each magnitude bin from the top panel.  There is a small
excess of $B \lesssim 19$  QSOs with absorbers compared with fainter
targets.  The solid line shows a least squares fit to these ratios 
(excluding the point at $B=17$) with a slope of $-0.3$.  The dashed
horizontal line shows a ratio of one.}
\end{figure}

In Figure \ref{mag_ratio} we reproduce an analogous plot to Figure 5
of MP03 and confirm qualitatively their result:
there is an excess of bright QSOs with absorbers.  Excluding
from our least squares fit the brightest magnitude bin 
which has a large error because of small number statistics, 
we find a slope of $-0.3$.  This gradient is somewhat shallower than
the value $-0.66$ found by MP03, although these authors also
show that the slope is  quite sensitive to the band selected.

In order to understand this dependence on magnitude, 
it is important to consider
the shape of the optical luminosity function (OLF) of QSOs.  If the
quasar OLF were a single power law, then absorber number statistics
should be independent of survey depth, \textit{even in the presence of 
dust or lensing bias}.  A bias due to dust or lensing
would alter the $n(z)$ determined, with a severity dictated by the
steepness of the LF, but one would not observe a varying
effect as a function of completeness magnitude.  However, at all
redshifts the quasar OLF is well modelled by a double power law with
a steep index at bright magnitudes and a considerably
shallower slope at fainter magnitudes, the turnover depending on
the QSO redshift e.g. (Boyle et al. 2000; although see
Hunt et al. 2004 for evidence of evolution in the faint end slope
at high redshift).  It is the transition between `bright' and `faint'
QSOs that causes the magnitude dependent effect seen in Figure  \ref{mag_ratio}
(and in Figure 5 of MP03).

The negative slope of the fit in Figure  \ref{mag_ratio} indicates
that gravitational lensing bias dominates over extinction effects
for bright QSOs with intermediate redshift absorbers,
as predicted by Perna et al. (1997). However, the wavelength
dependent nature of this effect, 
which is more apparent in red filters (MP03),
indicates that dust also makes a contribution.
In theory, we can disentangle the effects of dust and lensing by
looking at the ratio of QSOs with and without intervening absorbers
(as shown in Figure \ref{mag_ratio}) as a function of radio luminosity,
since lensing is achromatic, but dust extinction is not.
However, unlike the OLF, the quasar radio luminosity function for
flat spectrum sources at $2<z<3$ does not show a convincing break
(J. Wall, 2004, private communication). 
We therefore do not expect to find (and indeed do not find) a smooth
variation in the ratio of QSO numbers with/without intervening
absorbers as a function of radio
power.

\begin{figure}
\centerline{\rotatebox{0}{\resizebox{12cm}{!}
{\includegraphics{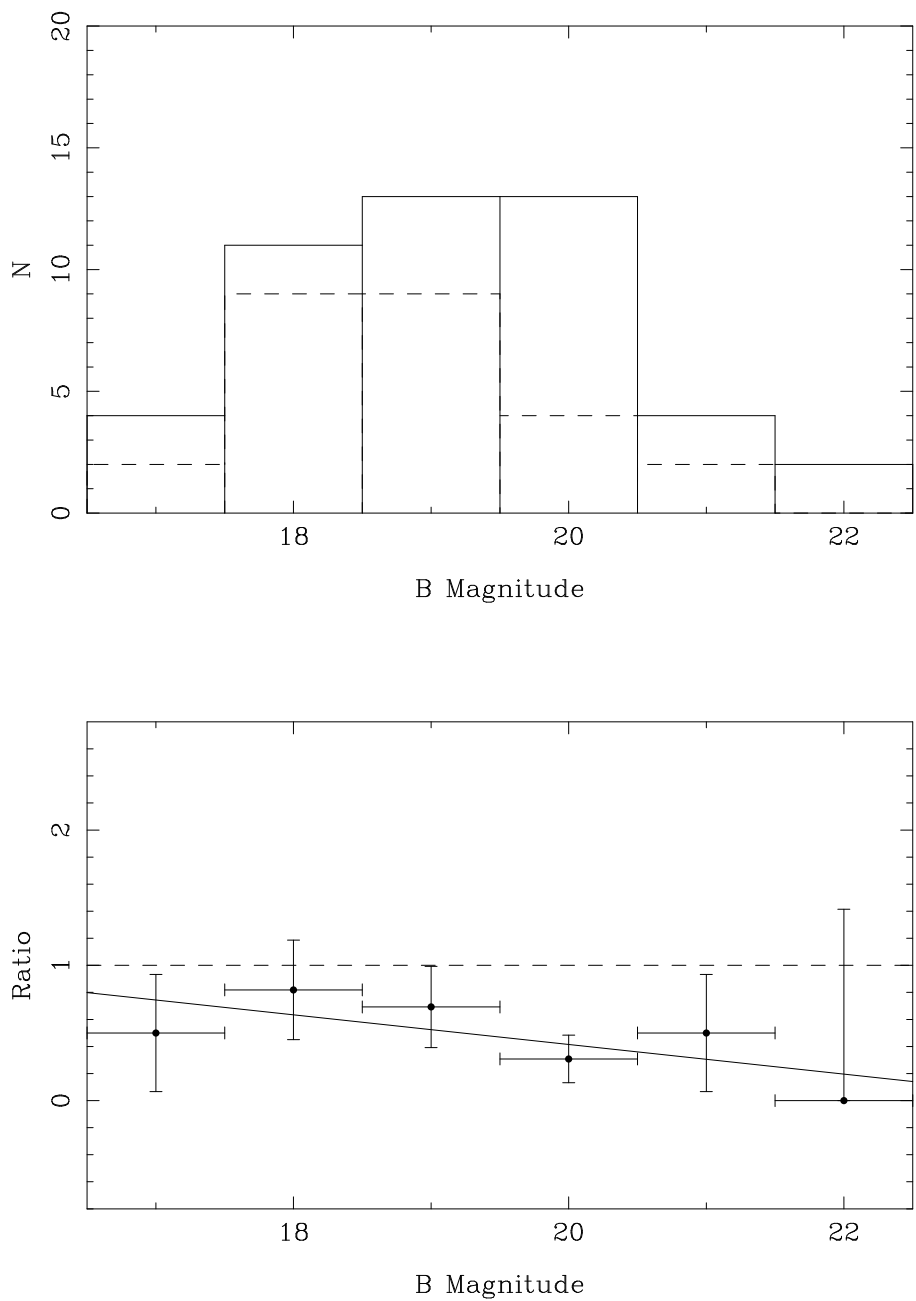}}}}
\caption{\label{mag_ratio2}  Top:  $B$-band magnitude distributions for QSOs
with (dashed line) and without (solid line) intervening \ion{Mg}{2} systems
with EW $\ge$ 0.6 \AA.  Bottom:  The ratio of the number of QSOs 
with:without intervening systems in each magnitude bin from the top panel.    
The solid line shows a least squares fit to these ratios 
with a slope of $-0.1$ and the dashed horizontal line shows a ratio of one.
This figure is analogous to Figure \ref{mag_ratio} except that the EW
cut-off ensures completeness in our absorption system selection.}
\end{figure}

An alternative explanation for the effect seen in Figure
\ref{mag_ratio} is that higher S/N in the spectra of brighter QSOs
facilitates the detection of absorbers towards these sources.  In order
to test this possibility we have compared the ratio of QSOs with
and without intervening absorbers with $EW  \ge 0.6$ \AA.  Since
we are effectively complete at this EW limit (see Figure \ref{gz})
imposing this cut should remove any bias due to S/N.  The results
of this test are shown in Figure \ref{mag_ratio2}, where it can
be seen that the `signature' of lensing bias seen in Figure \ref{mag_ratio}
vanishes.  Clearly, more investigation into gravitational lensing
bias with larger QSO samples and with careful quantification
of completeness is warranted.

We conclude this section by emphasizing that although the discussion
above acknowledges the possible existence of observational biases from dust
and lensing, both appear to be relatively minor effects.
Moreover, due to the change in the slope of the QSO OLF at $B \sim 19$,
surveys which have magnitude limits significantly fainter
than this fiducial value will not yield absorber statistics
which depend on QSO brightness.  In fact, the relatively shallow
OLF slope at faint magnitudes means that moderate dust (or
gravitational lensing bias) will have only a small effect and 
will not yield number densities significantly different from complete 
surveys such as CORALS.

\section{Discussion}

\subsection{DLA Statistics}\label{dla_sec}

The ultimate aim of this work is to quantify the obscuration bias
associated with DLAs, which form a subset of the strong \mgtwo\
absorbers identified here.
Although we can not make definitive statements concerning DLA
statistics until \textit{HST} observations are in hand, we can
speculate upon their number density based on the selection
statistics of previous surveys.
RT00 have shown that low redshift DLAs can
be efficiently pre-selected for UV follow-up based on the EWs of
strong metal lines.  
Specifically, they found in their sample of
87 \mgtwo\ systems that
50\% of absorbers with EW(\ion{Fe}{2} $\lambda$2600, \mgtwo $\lambda$2796) 
$>$ 0.5 \AA\ were confirmed to be DLAs based on \textit{HST} spectroscopy. 
Recently, P\'eroux et al (2004) have investigated
whether this pre-selection introduces a bias into DLA surveys, but
found no convincing evidence that this is the case.
In the absence of direct information on the \nhi\ of
our \mgtwo\ absorbers, we therefore adopt the RT00 pre-selection
as an indication of DLA statistics.

Our survey has been designed specifically with this pre-selection 
in mind.  In Figure \ref{gz_fe} we show the $g(z)$ of our sample
if we include the additional criterion of \ion{Fe}{2} $\lambda$2600
coverage.  Figure \ref{gz_fe} is therefore analogous to Figure \ref{gz}
except that the minimum EW thresholds now apply to both transitions.  
A comparison of the two figures shows that we achieve 
simultaneous coverage of these two metal lines for the majority of
sightlines.  There is a small ($\Delta z \sim 0.1$) gap in
coverage at the lowest redshift end which is unavoidable, simply because
the rest wavelength of \ion{Fe}{2} is bluer than that of \mgtwo.  There
is a second small dip in $g(z)$ at $1.3 < z_{\rm abs} < 1.4$ caused by 
the gap in CCD coverage in the ISIS instrument (there
is a corresponding gap for \mgtwo\ at $z_{\rm abs} \sim 1.18$).

In Figure \ref{dla_cand} we show the \mgtwo\ and \ion{Fe}{2} rest
frame EWs for every system in our catalogue (Table \ref{line_table}).  
In addition we plot the \ion{Mg}{1} $\lambda$2852 EW as a function
of \ion{Mg}{2} $\lambda$2796 EW; RT00 find that
all systems with EW(\ion{Mg}{1}$\lambda 2852) >$ 0.7 \AA\ are confirmed
to be DLAs.  Based on \mgtwo\ and \ion{Fe}{2} alone, we have 14
good DLA candidates, six of which have EW(\ion{Mg}{1} $\lambda 2852) >$ 0.7 
\AA.  According to the results of RT00, we would statistically expect
50\% of the 14 candidates to be confirmed as DLAs.  We can re-calculate
the redshift path incorporating \ion{Fe}{2} coverage for a given
sigma and EW limit in order to calculate $n(z)$ for the DLAs; these
values are given in Table \ref{dz_table}.

We have calculated $n_{\rm DLA}(z)$ for DLAs by assuming 7 absorption systems
in the range $0.6 < z_{\rm abs} < 1.6$  and $\Delta z$=44.62 (at 5 sigma 
significance for a 0.5 \AA\ detection), i.e. 
n(z)=0.16$^{+0.08}_{-0.06}$. 
The average absorption redshift of the 14 systems with 
EW(\ion{Mg}{2}$\lambda 2796$, \ion{Fe}{2}$\lambda 2600) >$ 0.5 \AA\ is
$\langle z \rangle = 1.07$.  In Figure \ref{nz_dla} we compare 
this value with previous DLA surveys
at high (Storrie-Lombardi \& Wolfe 2000) and low redshifts (RT00).  
We additionally include the $z=0$ number density from Rosenberg \& 
Schneider (2003)
which is in good agreement with other local \ion{H}{1} studies such as 
those by Ryan-Weber
et al. (2003) and Zwaan et al. (2002).  We conclude that the number density
of DLAs expected statistically in our sample is consistent with that
from previous surveys based on optically selected samples. However,
the small number of inferred systems (which needs to be confirmed
with UV spectroscopy) leads to large (1$\sigma$) error bars which
permit up to a factor of 2.5 difference with previous DLA surveys.  

\subsection{Average Metallicity of DLA Systems}

One of the outstanding questions concerning DLAs is why
their metallicities remain low ($\lesssim 1/10 Z_{\odot}$) , even at 
low-to-intermediate redshifts (e.g. Pettini et al. 1999).    
It has previously been
speculated (e.g. Pei \& Fall 1995; Pettini et al. 1999; Prantzos \& 
Boissier 2000;
Churches et al. 2004) that dust bias could cause an artificial
cut-off in the observed metallicities.  Akerman et al. (in preparation)
have found that this is not the case at high redshift: the mean
metallicity of $z>2$ DLAs from CORALS~I is in agreement with
that in optically selected samples.

We can attempt a crude comparison of metallicities between DLA
candidates in our survey and those in an optically selected sample,
namely the SDSS. Nestor et al. (2003b) have found that the mean 
metallicity  (based on [Zn/H] in composite spectra) of their SDSS
\mgtwo\ sample was higher for larger EW \mgtwo\ systems.
We therefore compare
the median \mgtwo\ EW from our survey with that determined from
the SDSS sample. Nestor et al (2003a) provisionally  fit the EW
distribution with an exponential of the form

\begin{equation}
n(EW) = n_0 e^{-EW/EW^{\star}}
\end{equation}

with a value of $EW^{\star}=0.7$ \AA.  This value is in very good agreement
with the previous determination of $EW^{\star}$ by SS92, who also
give an alternative formulation of

\begin{equation}\label{n_eqn}
n(EW) = C EW^{-\delta}
\end{equation}

with a value of $\delta=1.65$, which we adopt here.  
From Eq. (\ref{n_eqn}) we determine a median value of the EW
that depends on the minimum and maximum cut-offs adopted:

\begin{equation}\label{med_eqn}
2 EW_{median}^{1-\delta} =  EW_{max}^{1-\delta} + EW_{min}^{1-\delta}
\end{equation}

Applying to our CORALS~II sample the minimum EW cut-off
of Nestor et al. (2003b), EW=1 \AA, we 
measure a value of $EW_{median}$ = 1.50 \AA,
in good agreement with the theoretical 
median of 1.57 \AA\ determined from Eq. (\ref{med_eqn}), adopting the
maximum $EW = 3$~\AA\ found in our sample.  
This good correspondence between the median equivalent widths of the
CORALS~II and SDSS \mgtwo\ samples further suggests that the typical
metallicity of low redshift CORALS absorbers is unlikely 
to be significantly higher than that of absorbers from
optically selected QSO samples.  However, only a fraction
of DLAs will have sufficiently large \nhi\ and high metallicities
to cause severe obscuration; Cen et al. (2003) estimate a fraction
of $\sim$ 10\%.  Since our statistics are based on only 7 DLAs,
such a small sample of candidate DLAs is unlikely
to sample the full metallicity distribution function.

\section{Conclusions}

We have performed a new survey for \mgtwo\ absorption systems 
in the range $0.6 < z_{\rm abs} < 1.7$ from a radio
selected quasar sample with complete optical identifications.  
The strongest of these \mgtwo\ systems, particularly those with
accompanying \ion{Fe}{2} absorption, are good candidates for DLAs.  We have
used these data to quantify, for the first time, the effect of
dust bias on the completeness of optically selected surveys for low redshift
\mgtwo\ systems.

Our most robust result is that the number density of \mgtwo\ systems
is in excellent agreement with previous studies based on optically selected
QSOs.  Combined with the result of our high redshift survey (Ellison
et al. 2001), we have not yet found any evidence for a statistically
significant dust bias in absorption system surveys over the range 
$0.5 \lesssim z_{\rm abs} \lesssim 3.5$.  

At bright magnitudes ($ B \lesssim 19$) we observe a mild excess of QSOs
with intervening low redshift absorbers which may be due to
gravitational lensing bias.  However, this excess is not present
when only absorbers with $EW > 0.6$ \AA\ are included, indicating
that incompleteness may be at least partly responsible.  Our
discussion of gravitational lensing bias and its dependence on
the shape of the QSO OLF highlights an important conclusion for 
absorption line surveys: an optical completeness fainter the fiducial 
point in the OLF ($B \sim 19$) will not yield number densities
that are dependent on QSO magnitude.

Using the empirical pre-selection technique of Rao \& Turnshek (2000) we have 
attempted to estimate the number of DLAs in our sample and hence
their number density.  By assuming that 50\% of absorbers with
EW(\ion{Fe}{2} $\lambda$2600, \mgtwo $\lambda$2796) 
$>$ 0.5 \AA\ are statistically likely to be confirmed as DLAs, 
we determine the DLA number density at $\langle z \rangle = 1.07$ to be
$n(z)=0.16^{+0.08}_{-0.06}$.  This
is consistent with previous estimates of $n(z)$ at this redshift,
but the large 1$\sigma$ error bars permit up to a factor of 2.5
difference.  The number density relies both on an empirical
pre-selection and relatively small number statistics and should therefore
be confirmed with follow-up observations of the \lya\ transition.

Finally, the median equivalent width of the CORALS~II \mgtwo\ systems 
is in good agreement with that of optically selected absorbers
in the SDSS.  This is suggestive
of consistent metallicities between the two samples, since Nestor
et al. (2003b) find that larger EW \mgtwo\ systems have higher 
metallicities in their SDSS composites.

The next important step is to measure the \nhi\ of
our candidate DLAs.  This will permit us to calculate the neutral
gas mass density of our complete low redshift sample and allow
us to determine metallicities with ground-based telescopes.
Once we have determined the \nhi\ of these systems, there are
several foreseeable follow-up projects, such 
as 21~cm absorption measurements which will yield spin temperatures for
the absorbing galaxies.  The few measurements of spin
temperature that have already been obtained hint at some intriguing
trends (e.g. Kanekar \& Chengalur 2003), but there is an absence 
of data points in the
important range $0.7 < z < 1.7$.  Such follow-up programs will build
on the uniqueness of CORALS and exploit fully the benefits of this 
radio-selected
sample.

\acknowledgements

We are extremely grateful to Remi Cabanac who performed the March 2001
EFOSC2 run for us and to Paul Green and John Silverman for the opportunity
to obtain several spectra at the Baade telescope at Las Campanas Observatory.  
We thank Dan Nestor for providing the SDSS determinations
of $n(z)$ in advance of publication and Arif Babul and Jon Willis
for useful discussions on gravitational lensing.  SAR acknowledges
PPARC for a PhD studentship.
This work made use of the NASA Extragalactic Database (NED).




\begin{deluxetable}{cccccc}
\tablewidth{7.0in}
\tablecaption{\label{obs_table}Observing Journal} 
\tablehead{
\colhead{Telescope} & 
\colhead{Instrument} & 
\colhead{Grating/} & 
\colhead{Resolution} & 
\colhead{Wavelength } &
\colhead{Observing} \\
\colhead{} &
\colhead{} &
\colhead{Grism} &
\colhead{FWHM (\AA)} &
\colhead{Coverage (\AA)} &
\colhead{Dates} 
}
\startdata
3.6-m & EFOSC2 & \# 9 & 5.5  & 4730 -- 6700  & March 12-13 2002\\
3.6-m & EFOSC2 & \# 9 & 5.5  & 4730 -- 6700  & Nov. 9-11 2002 \\
WHT & ISIS & R600B & 1.9  & 4430 -- 6000  & March 17-18 2002\\
WHT & ISIS & R600R & 1.8  & 6185 -- 6970  & March 17-18 2002\\
WHT & ISIS & R600B & 1.9  & 4430 -- 6000   & Sept. 30 -- Oct 3 2002\\
WHT & ISIS & R600R & 1.8  & 6120 -- 7500\tablenotemark{a}   & Sept. 30 -- Oct 3 2002\\
UT3 & FORS1 & 600V & 5.1 & 4830 -- 7200   & Aug. 4-5 2002 \\ 
Baade & B\&C & 600/5000 & 4.1 & 4140 -- 7300  & Jan. 26 2003 \\ 
UT1 & FORS1 & 600V & 4.9 & 4830 -- 7200   & Mar--Aug 2003 \\ 
\enddata
\tablenotetext{a}{Unvignetted part of spectrum}
\end{deluxetable}


\begin{deluxetable}{lccccccc}
\tablewidth{7.0in}
\tablecaption{\label{qso_table}Target List, Exposure Times and QSO Redshift Coverage} 
\tablehead{
\colhead{QSO} & 
\colhead{$z_{\rm em}$} & 
\colhead{$B$ mag.\tablenotemark{a}} &  
\colhead{Telescope} & 
\colhead{Exposure} & 
\colhead{$\Delta z > 0.3$\AA\tablenotemark{b}} & 
\colhead{$\Delta z > 0.6$\AA\tablenotemark{b}} & 
\colhead{$\Delta z > 1.0$\AA\tablenotemark{b}} \\
\colhead{} & 
\colhead{} & 
\colhead{} &  
\colhead{} & 
\colhead{Time (s)} & 
\colhead{} & 
\colhead{} & 
\colhead{}
}
\startdata
   B0039-407 & 2.478 &  19.7 &  3.6  &  2700  & 0.0114& 0.5494& 0.7135 \\ 
   B0048-071 & 1.975 &  21.4 &  WHT  & 10800  & 1.0805& 1.1602& 1.1721 \\ 
   B0104-275 & 2.492 &  19.3 &  3.6  &  2700  & 0.0242& 0.7134& 0.7155 \\ 
   B0106+013 & 2.094 &  18.6 &  WHT  &  3000  & 1.0574& 1.1593& 1.1671 \\ 
   B0122-005 & 2.280 &  18.5 &  WHT  &  3600  & 1.1444& 1.1722& 1.1750 \\ 
   B0136-059 & 2.004 &  21.4 &  VLT  &  4000  & 0.0000& 0.2306& 0.8376 \\ 
   B0136-231 & 1.893 &  18.3 &  3.6  &  2000  & 0.7135& 0.7171& 0.7171 \\ 
   B0226-038 & 2.064 &  17.9 &  WHT  &  3600  & 1.1678& 1.1760& 1.1763 \\ 
   B0227-369 & 2.115 &  19.6 &  3.6  &  5400  & 0.0000& 0.0810& 0.7120 \\ 
   B0234-301 & 2.102 &  18.1 &  3.6  &  1800  & 0.0171& 0.7134& 0.7155 \\ 
   B0240-060 & 1.805 &  18.7 &  WHT  &  3600  & 1.1356& 1.1665& 1.1719 \\ 
   B0244-128 & 2.201 &  18.4 &  WHT  &  3600  & 0.6603& 1.1520& 1.1668 \\ 
   B0254-334 & 1.915 &  19.2 &  3.6  &  3600  & 0.0000& 0.2637& 0.7143 \\ 
   B0256-005 & 1.998 &  17.3 &  WHT  &  2800  & 1.1112& 1.1657& 1.1719 \\ 
   B0325-222 & 2.220 &  19.0 &  3.6  &  1800  & 0.0000& 0.2445& 0.7107 \\ 
   B0420+022 & 2.277 &  20.2 &  WHT  & 14400  & 1.1410& 1.1688& 1.1738 \\ 
   B0421+019 & 2.048 &  17.3 &  WHT  &  3600  & 1.1429& 1.1710& 1.1751 \\ 
   B0422-389 & 2.346 &  18.4 &  3.6  &  2700  & 0.0227& 0.7099& 0.7163 \\ 
   B0436-203 & 2.146 &  22.8$^\dagger$ &\nodata&\nodata & \nodata &\nodata&\nodata \\ 
   B0440-285 & 1.952 &  18.4 &  3.6  &  2700  & 0.5991& 0.7150& 0.7171 \\ 
   B0446-212 & 1.971 &  18.7 &  3.6  &  2400  & 0.0057& 0.7141& 0.7170 \\ 
   B0448-187 & 2.050 &  20.2 &  Baade  &  3600  & 0.8149& 1.1251& 1.1351  \\ 
   B0458-020 & 2.286 &  19.0 &  WHT  &  5400  & 0.2295& 1.0163& 1.1430 \\ 
   B0524-433 & 2.164 &  18.0$^\dagger$ &  3.6  &  3600  & 0.0078& 0.0953& 0.6995 \\ 
   B0606-223 & 1.926 &  20.0$^\dagger$ &  Baade  &  6000  & 1.1350& 1.1350& 1.1350 \\ 
   B0618-252 & 1.900 &  18.2$^\dagger$ &  3.6  &  3600  & 0.7035& 0.7170& 0.7170 \\ 
   B0642-349 & 2.165 &  18.5$^\dagger$ &  3.6  &  2600  & 0.3439& 0.7106& 0.7141 \\ 
   B0805-077 & 1.837 &  19.0$^\dagger$ &  WHT  &  3600  & 0.8451& 0.8458& 0.8458 \\ 
   B0819-032 & 2.352 &  19.4 &  WHT  &  7200  & 0.1699& 0.8352& 0.8420 \\ 
   B0919-260 & 2.300 &  19.0$^\dagger$ &  3.6  &  3600  & 0.7056& 0.7141& 0.7141 \\ 
   B0945-321 & 2.140 &  19.0$^\dagger$ &  3.6  &  3600  & 0.0000& 0.7024& 0.7116 \\ 
   B1005-333 & 1.837 &  18.0$^\dagger$ &  3.6  &  1800  & 0.0000& 0.5197& 0.7113 \\ 
   B1022-102 & 2.000 &  16.8$^\dagger$ &  WHT  &  3600  & 0.8432& 0.8459& 0.8459 \\ 
   B1032-199 & 2.198 &  18.2 &  WHT  &  5400  & 0.2719& 0.8377& 0.8412 \\ 
   B1034-374 & 1.821 &  18.5$^\dagger$ &  3.6  &  2600  & 0.0000& 0.6351& 0.7070 \\ 
   B1055-301 & 2.523 &  19.3 &  Baade  &  3600  & 1.0787& 1.1350& 1.1350 \\ 
   B1106-227 & 1.875 &  20.3 &  Baade  &  4500  & 1.0173& 1.1333& 1.1350 \\ 
   B1117-270 & 1.881 &  19.0$^\dagger$ &  3.6  &  3600  & 0.0590& 0.7088& 0.7131 \\ 
   B1143-245 & 1.940 &  18.1 &  3.6  &  1800  & 0.5703& 0.7120& 0.7141 \\ 
   B1147-192 & 2.489 &  20.3 &  Baade  &  3500  & 0.0245& 0.3658& 1.0166 \\ 
   B1148-001 & 1.978 &  17.4 &  WHT  &  2400  & 0.8433& 0.8459& 0.8459 \\ 
   B1149-084 & 2.370 &  20.0 &  VLT  &  1800  & 0.8377& 0.8465& 0.8485 \\ 
   B1228-310 & 2.276 &  19.8 &  3.6  &  5400  & 0.0000& 0.5197& 0.7056 \\ 
   B1230-101 & 2.394 &  19.6 &  Baade  &  2700  & 0.0446& 1.0363& 1.1344 \\ 
   B1255-316 & 1.924 &  18.3 &  3.6  &  2600  & 0.2506& 0.7127& 0.7141 \\ 
   B1256-177 & 1.956 &  21.4 &  VLT  &  1800  & 0.8464& 0.8480& 0.8480 \\ 
   B1256-243 & 2.263 &  19.4 &  3.6  &  5400  & 0.6277& 0.7131& 0.7131 \\ 
   B1318-263 & 2.027 &  21.3 &  VLT  &  4000  & 0.4884& 0.8393& 0.8439 \\ 
   B1319-093 & 1.864 &  19.6 &  WHT  &  9000  & 0.6455& 0.8396& 0.8426 \\ 
   B1324-047 & 1.882 &  19.8 &  WHT  &  9000  & 0.1574& 0.8366& 0.8407 \\ 
   B1402-012 & 2.518 &  18.0 &  WHT  &  2400  & 0.8404& 0.8457& 0.8459 \\ 
   B1406-267 & 2.430 &  21.8$^\dagger$ &  VLT  &  7200  & 0.7351& 0.8435& 0.8485 \\ 
   B1412-096 & 2.001 &  17.4 &  WHT  &  6000  & 0.2278& 0.8328& 0.8420 \\ 
   B1422-250 & 1.884 &  19.6$^\dagger$ &  3.6  &  5400  & 0.0178& 0.7017& 0.7095 \\ 
   B1430-178 & 2.331 &  19.4 &  3.6  &  3600  & 0.1905& 0.7131& 0.7131 \\ 
   B1451-400 & 1.810 &  18.5$^\dagger$ &  3.6  &  2600  & 0.0000& 0.3980& 0.7113 \\ 
   B1550-269 & 2.145 &  21.0$^\dagger$ &  VLT  &  3000  & 0.8404& 0.8470& 0.8487 \\ 
   B1654-020 & 2.000 &  23.5$^\dagger$ &\nodata&\nodata & \nodata &\nodata&\nodata \\ 
   B1657+022 & 2.039 &  19.2 &  WHT  &  6900  & 0.8126& 0.8430& 0.8449 \\ 
   B2044-168 & 1.943 &  18.0 &  WHT  &  3600  & 1.1667& 1.1752& 1.1755 \\ 
   B2123-015 & 2.196 &  19.9$^\dagger$ &  VLT  &  5000  & 0.0075& 0.7955& 0.8391 \\ 
   B2134+004 & 1.936 &  17.2 &  WHT  &  3600  & 1.1651& 1.1727& 1.1746 \\ 
   B2145-176 & 2.130 &  20.1 &  VLT  &  1600  & 0.0981& 0.8395& 0.8445 \\ 
   B2149-307 & 2.345 &  18.4 &  3.6  &  2700  & 0.7155& 0.7170& 0.7170 \\ 
   B2200-238 & 2.118 &  18.0 &  3.6  &  1800  & 0.2153& 0.7171& 0.7171 \\ 
   B2210-257 & 1.833 &  18.5 &  3.6  &  2700  & 0.0000& 0.7134& 0.7163 \\ 
   B2217-011 & 1.878 &  20.4$^\dagger$ &  VLT  &  5400  & 0.0278& 0.8362& 0.8437 \\ 
   B2224+006 & 2.248 &  22.0 &  VLT  &  7200  & 0.4198& 0.8424& 0.8466 \\ 
   B2245-128 & 1.892 &  18.3 &  WHT  &  3600  & 1.0310& 1.1588& 1.1686 \\ 
   B2245-328 & 2.268 &  18.3 &  3.6  &  2700  & 0.0590& 0.7150& 0.7171 \\ 
   B2254+024 & 2.090 &  17.8 &  WHT  &  3600  & 1.1570& 1.1718& 1.1749 \\ 
   B2311-373 & 2.476 &  18.4 &  3.6  &  2700  & 0.0732& 0.7135& 0.7171 \\ 
   B2314-409 & 2.448 &  19.0 &  3.6  &  1950  & 0.4775& 0.7155& 0.7170 \\ 
   B2315-172 & 2.462 &  21.0 &  VLT  &  2500  & 0.0133& 0.1155& 0.8424 \\ 
   B2325-150 & 2.465 &  20.0 &  VLT  &  1200  & 0.2600& 0.8474& 0.8482 \\ 
\enddata
\tablenotetext{a}{$B$ band magnitudes from the APM catalogue (accurate
to approximately $0.3$ magnitudes), except those
marked with a $\dagger$  which are from Jackson et al. (2002)}
\tablenotetext{b}{Redshift coverage based on a given EW detection
threshold at 5$\sigma$ significance.}
\end{deluxetable}


\begin{deluxetable}{lccc}
\tablewidth{4.5in}
\tablecaption{\label{line_table}\mgtwo\ $\lambda 2796$ and \ion{Fe}{2} 
$\lambda 2600$ Absorption Line Catalogue and Rest Frame Equivalent Widths} 
\tablehead{
\colhead{QSO} & 
\colhead{$z_{\rm abs}$} &
\colhead{$W_0$(Mg~II~$\lambda 2796$)} & 
\colhead{$W_0$(Fe~II~$\lambda 2600$)} \\
\colhead{} &
\colhead{} &
\colhead{ (\AA)} &
\colhead{ (\AA)\tablenotemark{a}} 
}
\startdata
B0039-407 & 0.8483 & 2.35 $\pm$ 0.15 & 2.09 $\pm$ 0.22 \\
B0048-071 & 1.4919 & 0.90 $\pm$ 0.03 & 0.36 $\pm$ 0.04 \\
B0048-071 & 1.5698 & 0.32 $\pm$ 0.03 & $<$0.07 \\
B0106+013 & 1.4256 & 0.47 $\pm$ 0.04 & 0.22 $\pm$ 0.05 \\
B0122-005 & 0.9949 & 1.56 $\pm$ 2.08 & 0.56 $\pm$ 0.38 \\
B0122-005 & 0.9973 & 0.23 $\pm$ 0.05 & $<$0.07 \\
B0136-231 & 0.8020 & 0.44 $\pm$ 0.06 & \nodata \\
B0136-231 & 1.1832 & 0.68 $\pm$ 0.06 & 0.27 $\pm$ 0.07 \\
B0136-231 & 1.2937 & 0.75 $\pm$ 0.06 & 0.23 $\pm$ 0.06 \\
B0226-038 & 1.3284 & 0.70 $\pm$ 0.02 & $<$0.37 \\
B0227-369 & 1.0289 & 0.59 $\pm$ 0.18 & 0.61 $\pm$ 0.16 \\
B0234-301 & 0.8238 & 0.91 $\pm$ 0.11 & 0.27 $\pm$ 0.20 \\
B0240-060 & 0.5810 & 1.44 $\pm$ 0.08 & \nodata \\
B0240-060 & 0.7550 & 1.65 $\pm$ 0.04 & 1.25 $\pm$ 0.04 \\
B0240-060 & 1.6310 & 0.34 $\pm$ 0.03 & $<$0.06 \\
B0244-128 & 0.8282 & 1.77 $\pm$ 0.09 & 1.23 $\pm$ 0.09 \\
B0244-128 & 1.2215 & 0.57 $\pm$ 0.11 & 0.36 $\pm$ 0.10 \\
B0254-334 & 1.1192 & 0.81 $\pm$ 0.19 & $<$0.25 \\
B0256-005 & 1.3369 & 1.70 $\pm$ 0.43 & 0.31 $\pm$ 0.12 \\
B0256-005 & 1.6134 & 0.25 $\pm$ 0.04 & $<$0.07 \\
B0420+022 & 0.9490 & 0.25 $\pm$ 0.05 & $<$0.07 \\
B0421+019 & 0.7394 & 0.48 $\pm$ 0.09 & $<$0.11 \\
B0421+019 & 1.6379 & 0.28 $\pm$ 0.03 & $<$0.07 \\
B0422-389 & 1.2956 & 0.62 $\pm$ 0.11 & $<$0.28 \\
B0458-020 & 0.8904 & 0.65 $\pm$ 0.12 & $<$0.27 \\
B0458-020 & 1.5271 & 2.09 $\pm$ 0.82 & 0.44 $\pm$ 0.13 \\
B0458-020 & 1.5605 & 0.94 $\pm$ 0.08 & 0.75 $\pm$ 0.08 \\
B0606-223 & 0.5078 & 0.25 $\pm$ 0.04 & \nodata \\
B0606-223 & 0.8959 & 0.55 $\pm$ 0.02 & 0.41 $\pm$ 0.03 \\
B0606-223 & 0.9343 & 0.23 $\pm$ 0.02 & $<$0.06 \\
B0606-223 & 1.2443 & 1.40 $\pm$ 0.04 & 1.07 $\pm$ 0.04 \\
B0606-223 & 1.5313 & 1.55 $\pm$ 0.07 & 1.27 $\pm$ 0.05 \\
B0919-260 & 0.7048 & 0.81 $\pm$ 0.08 & \nodata \\
B0919-260 & 0.7626 & 0.35 $\pm$ 0.07 & \nodata \\
B0945-321 & 1.1702 & 0.91 $\pm$ 0.12 & $<$0.26 \\
B1005-333 & 1.3734 & 0.93 $\pm$ 0.10 & 0.84 $\pm$ 0.11 \\
B1022-102 & 0.7141 & 0.34 $\pm$ 0.03 & $<$0.08 \\
B1022-102 & 1.3085 & 0.17 $\pm$ 0.03 & \nodata \\
B1148-001 & 1.2450 & 0.31 $\pm$ 0.05 & $<$0.11 \\
B1148-001 & 1.4671 & 0.22 $\pm$ 0.03 & 0.10 $\pm$ 0.04 \\
B1230-101 & 0.7472 & 2.10 $\pm$ 0.15 & $<$1.33 \\
B1230-101 & 0.7810 & 0.86 $\pm$ 0.10 & $<$0.26 \\
B1255-316 & 0.6923 & 0.34 $\pm$ 0.10 & \nodata \\
B1256-177 & 0.9399 & 2.96 $\pm$ 0.04 & 2.08 $\pm$ 0.04 \\
B1256-177 & 1.3667 & 1.28 $\pm$ 0.03 & 0.33 $\pm$ 0.03 \\
B1256-177 & 1.5035 & 0.26 $\pm$ 0.03 & $<$0.08  \\
B1318-263 & 1.1080 & 1.38 $\pm$ 0.08 & 0.61 $\pm$ 0.07 \\
B1324-047 & 0.7850 & 2.58 $\pm$ 0.12 & 1.77 $\pm$ 0.11 \\
B1402-012 & 0.8901 & 1.21 $\pm$ 0.04 & 0.99 $\pm$ 0.04 \\
B1412-096 & 1.3464 & 0.66 $\pm$ 0.11 & \nodata \\
B1430-178 & 1.3269 & 0.60 $\pm$ 0.08 & 0.45 $\pm$ 0.07 \\
B1451-400 & 0.9330 & 0.71 $\pm$ 0.18 & $<$0.42 \\
B2044-168 & 0.8341 & 0.23 $\pm$ 0.03 & $<$0.06 \\
B2044-168 & 1.3287 & 0.59 $\pm$ 0.02 & $<$0.14 \\
B2149-307 & 1.0904 & 1.45 $\pm$ 0.04 & 0.78 $\pm$ 0.04 \\
B2245-128 & 0.5869 & 1.28 $\pm$ 0.08 & \nodata \\
B2314-409 & 1.0439 & 0.39 $\pm$ 0.08 & $<$0.21 \\
\enddata
\tablenotetext{a}{`\nodata' indicates that the transition was not covered}
\end{deluxetable}


\begin{deluxetable}{lcccc}
\tablewidth{4.0in}
\tablecaption{\label{dz_table}Redshift Path for Complete QSO Sample for
Various Detection Thresholds} 
\tablehead{
\colhead{Coverage} &
\colhead{0.3 \AA} &
\colhead{0.5 \AA} & 
\colhead{0.6 \AA} &
\colhead{1.0 \AA} 
}
\startdata
3 $\sigma$ \mgtwo\ only & 54.05 & 63.38 & 63.77 & 64.19 \\
5 $\sigma$ \mgtwo\ only & 35.15 & 54.05 & 58.24 & 63.77 \\
3 $\sigma$ \mgtwo\ and \ion{Fe}{2} & 44.62 & 53.51 & 53.85 & 55.10 \\
5 $\sigma$ \mgtwo\ and \ion{Fe}{2} & 28.11 & 44.62 & 48.52 & 53.85 \\
\enddata
\end{deluxetable}


\begin{deluxetable}{cccccccccccc}
\tablewidth{7.0in}
\tablecaption{\label{stats}Number of Systems, Mean Redshift, and 
Redshift Path for the RT00, SS92 and CORALS Samples at 5$\sigma$ significance.}
\tablehead{ 
\colhead{} &
\multicolumn{3}{c}{RT00} &
\colhead{} &
\multicolumn{3}{c}{SS92} &
\colhead{} &
\multicolumn{3}{c}{CORALS~II} \\ [.2ex] \cline{2-4} \cline{6-8} \cline{10-12}
\colhead{$W_{min}$ (\AA)} & 
\colhead{$N_{MgII}$} &
\colhead{$<z>$} &
\colhead{$\Delta z$} &
\colhead{} &
\colhead{$N_{MgII}$} &
\colhead{$<z>$} &
\colhead{$\Delta z$}&
\colhead{} &
\colhead{$N_{MgII}$} &
\colhead{$<z>$} &
\colhead{$\Delta z$}
}
 
\startdata
0.3 & 87 & 0.83 & 104.6 &  & 111 & 1.12 & 114.2 &  & 45 & 1.09 & 35.15\\
0.6 & 44 & 0.83 & 103.7 &  & 67  & 1.17 & 129.0 &  & 32 & 1.09 & 58.24\\
1.0 & \nodata &\nodata& \nodata&  & \nodata&\nodata& \nodata&  & 17 & 1.01 & 63.77\\
\enddata
\end{deluxetable}

\begin{figure}
\centerline{\rotatebox{270}{\resizebox{12cm}{!}
{\includegraphics{f7.eps}}}}
\caption{\label{gz_fe}The redshift path density, $g(EW_{min},z_i)$ for three
minimum EW limits of both \mgtwo\ and \ion{Fe}{2}, as a function of redshift.  
Dotted, solid and dashed
lines are for EW limits of 1.0, 0.6 and 0.3~\AA\ (at 5 $\sigma$ significance) respectively. }
\end{figure}

\begin{figure}
\centerline{\resizebox{12cm}{!}
{\includegraphics{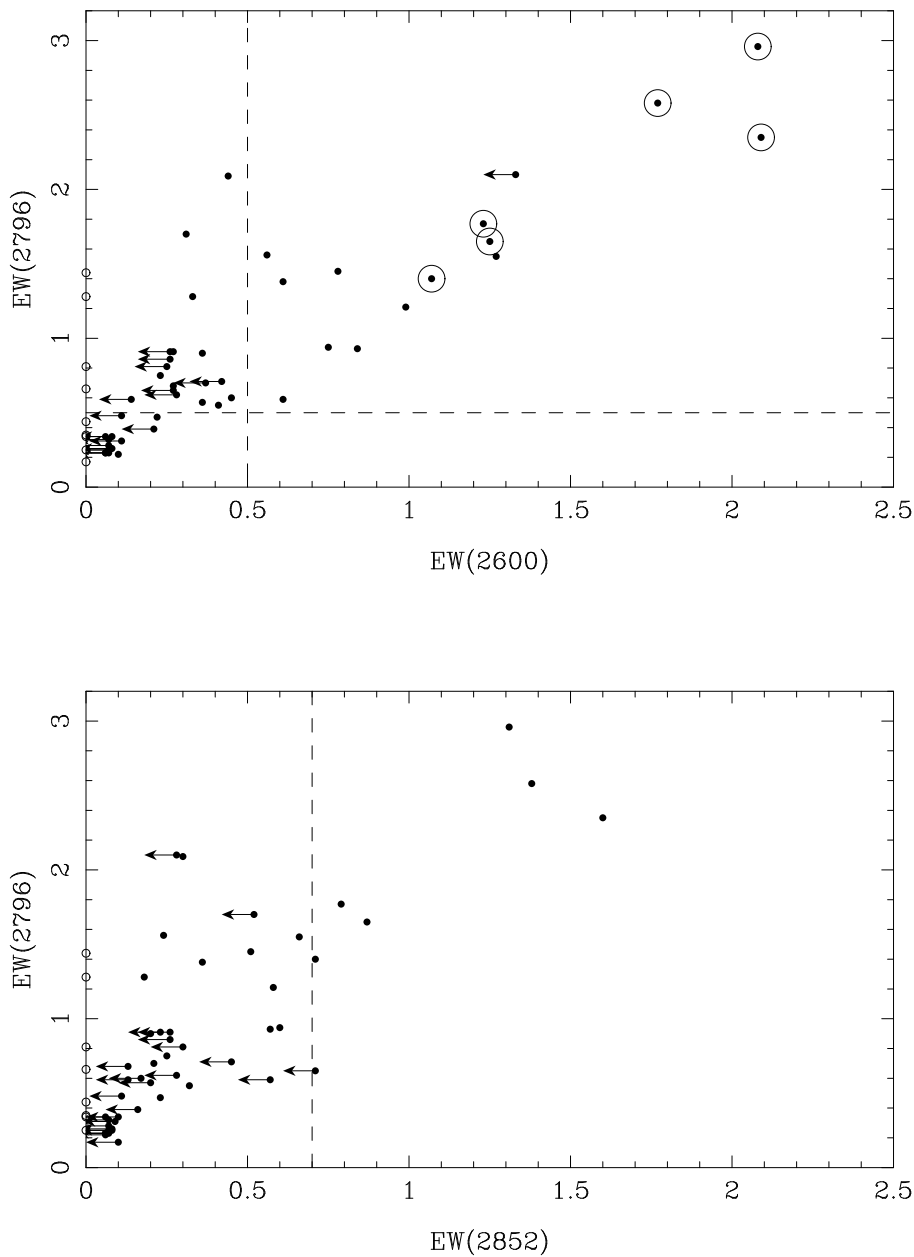}}}
\caption{\label{dla_cand} Top panel: Comparison of \ion{Mg}{2} and
\ion{Fe}{2} EWs.  DLA candidates can be selected from absorbers
that have rest equivalent widths of \ion{Mg}{2}$\lambda$2796 and 
\ion{Fe}{2}$\lambda 2600$ greater than 0.5 \AA\ (dashed lines).  
Solid points surrounded
with large open circles are systems with EW(\ion{Mg}{1}
$\lambda 2852) >$ 0.7 \AA. Small open points along
the y axis indicate that \ion{Fe}{2}$\lambda 2600$ was not covered 
in the spectrum.  Bottom panel:  Comparison of
\mgtwo\ and \ion{Mg}{1} EWs; symbols as for upper panel.  RT00 found
that all systems with EW(\ion{Mg}{1}$\lambda 2852)>0.7$ \AA\ were
confirmed to be DLAs (dashed line).}
\end{figure}

\begin{figure}
\centerline{\rotatebox{270}{\resizebox{12cm}{!}
{\includegraphics{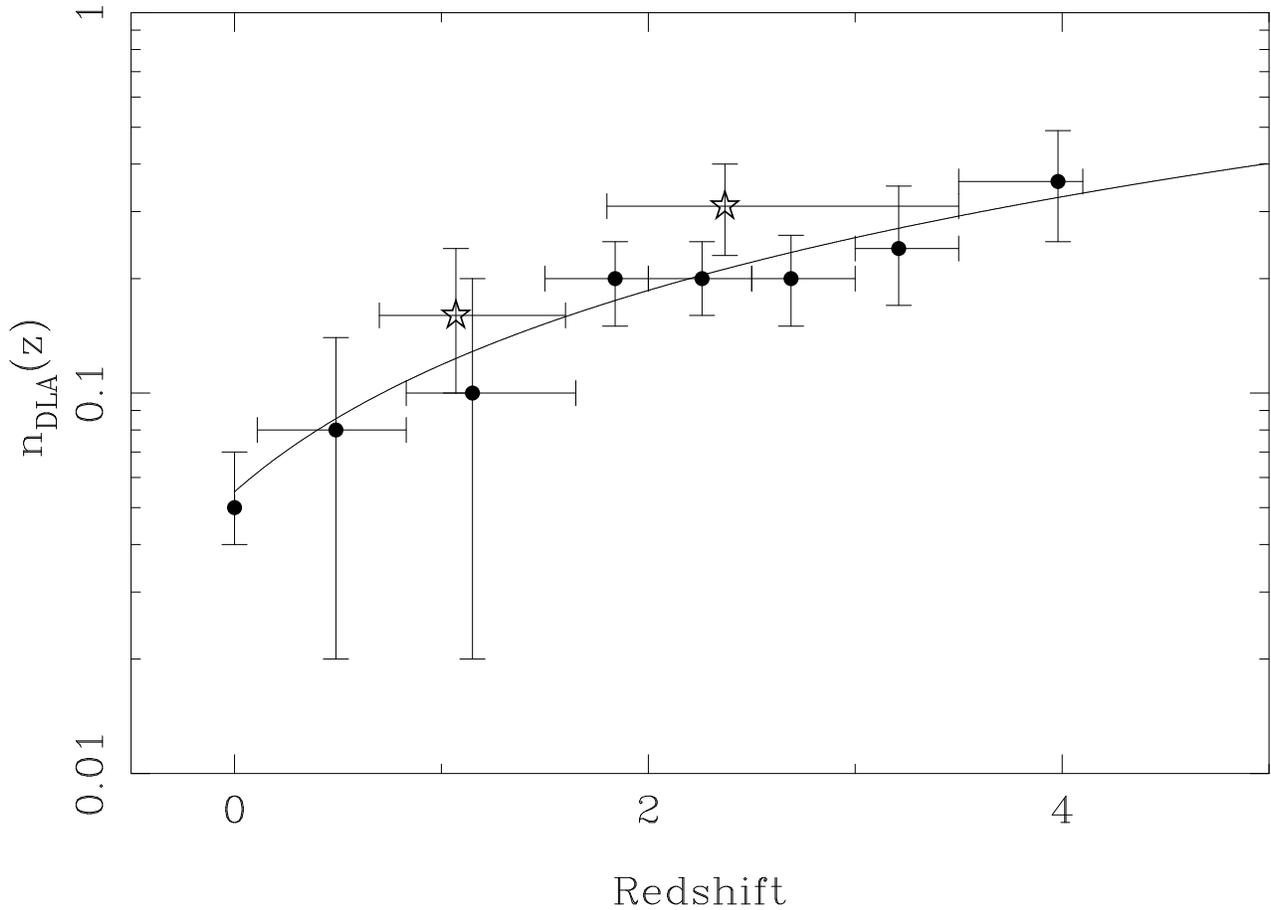}}}}
\caption{\label{nz_dla} The number density of DLAs as a function of redshift.
Solid points taken from the literature are from Rosenberg \& Schneider (2003),
Rao \& Turnshek (2000) and Storrie-Lombardi \& Wolfe (2000) for z=0, 
$0.1<z<1.65$ and $z>1.5$ respectively. .  The open
stars are the values inferred for the high (Ellison et al 2001) and
low (this work) redshift CORALS surveys. All error bars are 1 $\sigma$ based
on Gehrels (1986) except for the $z=0$ point where the error is
as quoted by Rosenberg \& Schneider (2003).  The solid line shows the
fit of Storrie-Lombardi \& Wolfe (2000): $n(z)=0.055(1+z)^{1.11}$.}
\end{figure}

\end{document}